\documentclass[%
aip,
amsmath,amssymb,
tightenlines,
floatfix,
reprint
]{revtex4-1}

\usepackage{graphicx}
\usepackage{dcolumn}
\usepackage{bm}

\usepackage[utf8]{inputenc}
\usepackage[T1]{fontenc}

\usepackage{newtxtext,newtxmath}
\usepackage{microtype}

\usepackage{booktabs}
\usepackage{mathtools}
\usepackage{braket}

\usepackage{makecell}

\usepackage[cal=dutchcal]{mathalpha}

\usepackage[svgnames]{xcolor}
\definecolor{cset-aps-blueberry}{RGB}{28,128,158}
\definecolor{cset-aps-blue}{RGB}{46,44,184}
\definecolor{cset-aps-turquoise}{RGB}{0,67,88}
\definecolor{cset-aps-limegreen}{RGB}{190,219,67}
\definecolor{cset-aps-green}{RGB}{31,138,112}
\definecolor{cset-aps-yellow}{RGB}{255,225,25}
\definecolor{cset-aps-orange}{RGB}{253,116,0}
\definecolor{cset-aps-red}{RGB}{219,0,43}

\definecolor{cset-aps-kobalt-medium}{RGB}{62,54,222}
\definecolor{cset-aps-kobalt-dark}{RGB}{28,24,150}

\definecolor{cset-aps-my-label-red}{RGB}{202,0,17}
\definecolor{cset-aps-my-label-blue}{RGB}{53,71,140}
\definecolor{cset-aps-my-label-gray}{RGB}{145,145,145}

\usepackage{hyperref}
\hypersetup{%
    colorlinks=true,
    linkcolor={cset-aps-red},
    linkbordercolor={cset-aps-red},
    filecolor={cset-aps-orange},
    filebordercolor={cset-aps-orange},
    citecolor={cset-aps-kobalt-medium},
    citebordercolor={cset-aps-kobalt-medium},
    urlcolor={cset-aps-kobalt-dark},
    urlbordercolor={cset-aps-kobalt-dark},
    menucolor={cset-aps-limegreen},
    menubordercolor={cset-aps-limegreen},
    breaklinks=true,
    pdfborderstyle={/S/U/W 2},
    pdfpagemode=UseOutlines,
    pdfstartpage={1},
}

\usepackage{nicefrac}

\newcommand{\orcid}[1]{\href{https://orcid.org/#1}{\includegraphics[width=7pt,height=7pt]{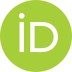}}}

\newcommand{\affULM}{\address{Institut f{\"u}r Quantenphysik and Center for Integrated Quantum Science and Technology (IQST), Universit{\"a}t Ulm, Albert-Einstein-Allee 11, D-89081 Ulm, Germany}}
\newcommand{\affTUDa}{\address{Technische Universit{\"a}t Darmstadt, Fachbereich Physik, Institut f{\"u}r Angewandte Physik, Schlossgartenstr. 7, D-64289 Darmstadt, Germany}}

\newcommand{\ii}{i}

\newcommand{\ie}{i.\,e.,}

\newcommand{\dd}{\text{d}}
\newcommand{\vect}[1]{\bm{#1}}

\DeclareMathOperator{\sinc}{sinc}

\begin{document}

\title{Unified laboratory-frame analysis of atomic gravitational-wave sensors}
\collaboration{This article has been published in \href{https://doi.org/10.1116/5.0304468}{AVS Quantum Sci. \textbf{7}, 044402 (2025)} as part of the Special Collection \href{https://pubs.aip.org/aqs/collection/619039/Advances-in-Matter-Wave-Optics}{\emph{Advances in Matter Wave Optics}} under the terms of the \href{https://creativecommons.org/licenses/by/4.0/}{Creative Commons Attribution License 4.0 [CC BY]}}

\preprint{AIP/123-QED}

\author{Simon Schaffrath\,\orcid{0009-0004-4829-6733}}
\email{simon-schaffrath@gmx.de}
\affTUDa
\author{Daniel Störk\,\orcid{0009-0007-4155-0665}}
\affULM
\author{Fabio Di Pumpo\,\orcid{0000-0002-6304-6183}}
\email{fabio.di-pumpo@uni-ulm.de, fabio.di-pumpo@gmx.de}
\affULM
\author{Enno Giese\,\orcid{0000-0002-1126-6352}\,}
\affTUDa

\begin{abstract}
Atomic sensors using light-matter interactions, in particular atomic clocks and atom interferometers, have the potential to complement optical gravitational-wave detectors in the mid-frequency regime.
Although both rely on interference, the interfering components of clocks are spatially colocated, whereas atom interferometers are based on spatial superpositions.
Both the electromagnetic fields that drive the transitions and generate superpositions, while propagating through spacetime, as well as the atoms themselves as massive particles are influenced by gravitational waves, leading to effective potentials that induce phase differences inferred by the sensor.
In this work, we analyze the effects of these potentials on atomic clocks and atom interferometers in the laboratory frame.
We show that spatial superpositions in atom interferometers, both light-pulse and guided ones, give rise to a gravitational-wave signal.
Although these spatial superpositions are suppressed for clocks, we show that the light pulses driving internal transitions measure the spatial distance between the centers of two separate clocks.
We highlight that this mechanism only yields a sensitivity if both clocks, including possible trapping setups, move on geodesics given by the gravitational wave.
While such configurations are natural for satellite free-fliers, terrestrial optical clocks usually rely on stationary traps, rendering them insensitive to leading order.
Moreover, we show that both sensors can be enhanced by composite interrogation protocols in a common framework.
To this end, we propose a pulse sequence that can be used for large-momentum-transfer atom interferometers and for hyper-echo atomic clocks, leading to a signal enhancement and noise suppression.
\end{abstract}

\maketitle

\section{\label{sec:Intro}Introduction}
So far, gravitational waves (GWs) have only been directly detected~\cite{abbott2016, abbott2016a} by optical interferometers, such as LIGO,~\cite{Abbott2009} VIRGO,~\cite{Acernese2015} or KAGRA.~\cite{Aso2013}
Taking this technology to the next level, future space missions such as LISA~\cite{danzmann1996,bayle2022} or terrestrial third-generation detectors such as the Einstein telescope~\cite{Punturo2010} are planned with increased sensitivities.
Moreover, the use of optical resonators based on cavities has been proposed.~\cite{Barontini2025}
Despite the success of these setups, atom-interferometric antennas~\cite{Dimopoulos2008Atomic_gravitational,Dimopoulos2009} consisting of atoms in spatial superposition, driven by the same light fields,~\cite{Yu2011} may complement these efforts, focusing on the mid-frequency regime.
To this end, specific atom-interferometer designs have been proposed,~\cite{Graham2013,Schubert2024} resulting in the planning and commissioning of terrestrial prototypes.~\cite{Canuel2018,Badurina2020,Zhan2020,Canuel2020,El-neaj2020,Abe2021,Abend2023,Abdalla2025}
Together with their optical counterparts, they will help to establish a network of GW detectors, ultimately aiming at the separation of astrophysical GW sources~\cite{Sathyaprakash2009,abbott2016} from cosmological ones.~\cite{Maggiore2000,Caprini2018}
However, atoms possess not only quantized spatial degrees of freedom, but also internal energy levels which can be used to build atomic clocks, in principle susceptible to GWs as well.~\cite{Armstrong2006,Vutha2015,Kolkowitz2016}

Hence, the two prominent atomic sensors for GW detectors, namely atomic clocks and atom interferometers, both make use of quantum superpositions in different ways.
Their sensitivities to GWs have been compared~\cite{Norcia2017} in freely falling frames, together with sensitivity-enhancement techniques based on large-momentum-transfer~\cite{Rudolph2020,Gebbe2021,Berg2015} and hyper-Ramsey or hyper-echo~\cite{Yudin2010,Zanon2015,Hobson2016} inspired schemes~\cite{Zanon2025} that rely on dynamical decoupling.~\cite{Viola1999,Suter16}
We complement the discussion by identifying the origin of the susceptibility of both types of sensors to GWs in the laboratory frame:
(i) In atom interferometers, the spatial superposition of two arms probes an oscillating gravitational potential caused by the GW and acting on the center-of-mass (c.m.) position of the atoms.
(ii) In contrast, in optical atomic clocks, a superposition of their c.m. degrees of freedom is suppressed by trapping atoms in the Lamb--Dicke regime.~\cite{Dicke1953,Ludlow2015}
However, the transitions necessary to generate internal superpositions probe the c.m. position of the atom. 
Consequently, an atomic clock can only be sensitive to GWs if its c.m. position follows a geodesic trajectory within a GW.
Hence, the atom and therefore the center of the trapping potential have to fall freely.
Otherwise, for stationary traps, which represent the usual choice~\cite{Paul1990,Zhukas2021,Martinez2022} in terrestrial optical clocks fixed to an optical table, the leading-order effect is a mere readout of a constant position, and the GW enters only as a suppressed effect.
As a consequence, trapped clocks are only sensitive to GWs if the trap moves on a geodesic, which is natural in space~\cite{Kolkowitz2016,Norcia2017,Ebisuzaki2020,Fedderke2022} but requires extremely challenging solutions for terrestrial setups, such as a suspension~\cite{Cumming2012,Matichard2015} comparable to optical detectors.
While terrestrial fountain clocks~\cite{Kasevich1989,Wynands2005} are not operated at optical frequencies, leading to a decreased sensitivity, freely falling clocks~\cite{DiPumpo2023} on Earth with suppressed momentum transfer~\cite{Alden2014,Janson2024} are in principle also possible, but suffer from the same decrease in sensitivity, besides relying on two-photon transitions that might introduce differential laser-phase noise.~\cite{Yu2011}
In contrast, due to their spatial c.m. superposition, atom interferometers do not rely on this mechanism of reading out the c.m. position through light pulses, and thus they can be both freely falling or guided.

While previous work often focused on specific couplings to GWs, we take into account all leading-order couplings through different mechanisms.
We include the oscillating gravitational potential acting on the c.m. position of the atom and coupling to the rest mass~\cite{Maggiore2007,Berlin2022,Badurina2025} as well as to the internal energy through the mass defect,~\cite{Zych2011,Yudin2018,Sonnleitner2018,Schwartz2019,Martinez2022,Asano2024} but also through the propagation of light in the GW~\cite{Misner1973,Tsagas2005,Mieling2021,Ruggiero2025}, modifying the light's frequency and its wave vector.
As such, we identify contributions which can be attributed to Doppler, Einstein, and Shapiro phase shifts.~\cite{Badurina2025}
Finally, we propose composite interrogation protocols~\cite{Graham2013,Norcia2017,Berg2015} which are in principle identical for atom interferometers and clocks, leading to a sensitivity enhancement through large-momentum-transfer and hyper-echo schemes, respectively.

We establish in Sec.~\ref{sec:AtomsLightGW} a branch-dependent description for atomic clocks and atom interferometers in a common framework, taking into account all coupling mechanisms to GWs.
Section~\ref{sec:Principles} highlights the origin of the GW sensitivity, explicitly showing the difference between both types of sensors and the need for a geodesic motion for clocks.
In Sec.~\ref{sec:Enhance} we introduce a common description for composite pulse sequences to enhance the signals of both clocks and atom interferometers.
Finally, Appendix~\ref{app.GW-potential} provides the derivation of potentials for the atom interacting with a GW as well as light pulses that propagated through the GW.
Appendix~\ref{app.classical_EOMs} presents classical equations of motion for clocks and their solutions, while explicit expressions for large-momentum-transfer and hyper-echo schemes are given in Appendix~\ref{app.composite_pulses}.

\section{\label{sec:AtomsLightGW}Atoms and light pulses in gravitational waves}
\begin{figure}
    \centering
    \includegraphics[width=\columnwidth]{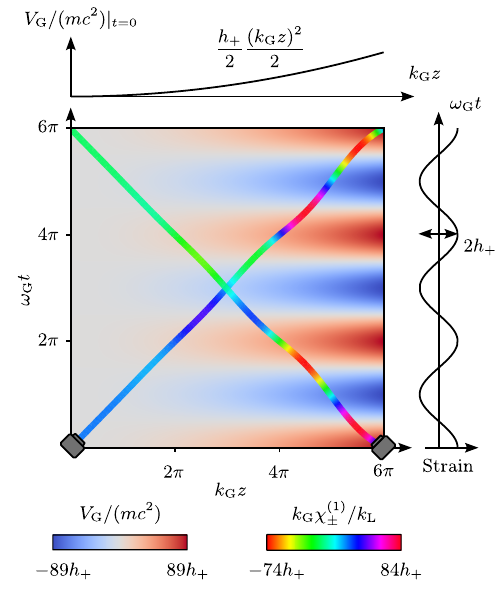}
    \caption{
    Gravitational potential $\hat{V}_\text{G}$ induced by a GW in the laboratory frame in space and time (density plot).
    The harmonic potential (shown on the top for $t=0$) oscillates between positive and negative values in phase with the strain of the GW (shown on the right).
    The light cones for pulses traveling in a GW background from the left ($+$) and right ($-$) are modified as well.
    The phase perturbation $\chi_\pm^{(1)}$ of the respective light beam is shown by the hue color scale along the light cone.
    If evaluated at the time and position where the light interacts with the atom, this perturbation translates into the potentials $\hat{V}_{\omega,\text{G}}$ and $\hat{V}_{k,\text{G}}$.
    }
    \label{fig:Potentials}
\end{figure}

In this section, we summarize the effects of a GW acting on a possibly trapped atom with mass $m$, which interacts with light pulses.
In the laboratory reference frame, it is described by the Hamiltonian
\begin{equation} 
\label{eq:hamiltonian}
\begin{split}
    \hat{H} = \frac{\hat{p}^2}{2 m} + \hat{V}_\text{T} + \hat{V}_\omega+ \hat{V}_k  + \hat{V}_\text{G}+ \hat{V}_{\omega,\text{G}}+\hat{V}_{k,\text{G}} + \hat{V}_{\Delta m,\text{G}},
\end{split}
\end{equation}
with $\hat{p}$ being the momentum operator.
We omit Earth's gravity and rotations to focus on GWs, as they do not modify our results to leading order, especially in differential measurements.

\begin{subequations}
To describe trapped optical clocks (or in principle also guided atom interferometers), we include the potential 
\begin{equation}
    \hat{V}_\text{T} = \frac{m \omega_\text{T}^2}{2} (\hat{z}-z_T)^2
\end{equation}
with trapping frequency $\omega_\text{T}$, which is zero for freely falling configurations, and centered around the position $z_T$ otherwise.
Here, the position $\hat{z}$ and momentum $\hat{p}$ operators fulfill the commutator relation $[\hat{z},\hat{p}]=\ii \hbar $.

We account for light pulses of frequency $\omega_\text{L}$ driving internal transitions between two atomic levels with energy splitting $\hbar\omega_\text{A}$ through
\begin{equation}
        \hat{V}_\omega=  \hbar \dot \Lambda_\omega (\omega_\text{L}-\omega_\text{A}) t,
\end{equation}
imprinting a phase difference between internal states and the light fields driving the transitions.
The pulsed nature is represented by time-dependent coefficients $\Lambda_\omega = \pm 1/2$ accounting for the atom being in the excited or ground state.
Consequently, $\dot \Lambda_\omega$ is a directed series of delta functions at the times of the pulses, where the sign is associated with excitation and deexcitation.
The particular sequence of pulses depends on the interferometer branch introduced later.
Note that $\dot \Lambda_\omega$ vanishes if no light pulses are applied, e.\,g., in some guided atom-interferometer setups.
We do not include finite speed of light on the scale of single sensors~\cite{Tan2016,DiPumpo2023,Niehof2025} but will introduce it between two sensors as a crucial ingredient~\cite{Dimopoulos2009,Yu2011} to infer the GW signal.

In addition to internal transitions, the absorption of photons induces a momentum transfer $\hbar k_\text{L} $ of the atom at the time of the pulse, where $k_\text{L}=\omega_\text{L}/c$ is the light's wave vector and $c$ the speed of light.
We define this transfer in $z$ direction, yielding the kick potential
\begin{equation}
    \hat{V}_k= - \hbar \dot \Lambda_k k_\text{L} \hat{z},
\end{equation}
resulting from the spatial dependence of the laser phase, and creating spatial superpositions, which are crucial for atom interferometers but suppressed in the Lamb--Dicke regime for atomic clocks.
Here, we define $\dot \Lambda_k$ as a series of directed delta pulses as before, but now the sign depends on the direction of the branch-dependent momentum transfer ($+$ up and $-$ down), which provides a possibility to account for large-momentum transfer.

All potentials introduced so far are necessary to describe the dominant contributions of the quantum sensors.
Additionally, we include all GW couplings as perturbative potentials.
Hence, we consider the potential created by a GW of frequency $\omega_\text{G} = c k_\text{G}$ and strain $h_+$ in $+$ polarization
\begin{equation}
    \hat{V}_\text{G} = \frac{h_+}{2} \frac{m \omega_\text{G}^2}{2} \hat{z}^2 \cos(\omega_\text{G}t + \varphi),
\end{equation}
\end{subequations}
acting on a massive particle in the laboratory frame derived in Appendix~\ref{app.GW-potential}.
Here, the $z$ direction is the transverse direction of the propagating GW with a phase offset $\varphi$.
It acts as a time-dependent harmonic gravitational potential for atoms, shown in the density plot in Fig.~\ref{fig:Potentials}.
The phase induced by this gravitational potential can be connected to the Doppler phase shift.~\cite{Badurina2025}
Hence, the GW has no effect for a particle located at the origin $z=0$. 
In fact, it corresponds to the point of expansion of our laboratory system, and in this reference frame all distances are measured with respect to this origin.
While in other frames the distances between two particles may be directly changed by the metric, in the laboratory frame the potential instead accelerates them, depending on their location.

At the same time, also the interrogating lasers propagate through the GW.
In Appendix~\ref{app.GW-potential} we solve the Eikonal equation for such a setting, resulting in a perturbation $\chi_{\pm}^{(1)} = \chi_\omega \pm \chi_k $ of the light's phase, shown as a density plot along the light cone in Fig.~\ref{fig:Potentials} for pulses traveling in positive (+) and negative (-) $z$ direction.
First, $\chi_\omega$ leads to an altered frequency of the laser pulse depending on the position, which generates the potential
\begin{subequations}
\begin{equation}
    \hat{V}_{\omega,\text{G}} = \hbar \frac{h_+}{2} \dot \Lambda_\omega   \frac{k_\text{L}}{k_\text{G}} \left( \frac{(k_\text{G} \hat{z})^2}{2}-1\right) \sin(\omega_\text{G}t + \varphi).
\end{equation}
The identification with the frequency part of the light field is possible because $\chi_\omega$ does not depend on the direction of the light pulse.
Second, also the spatial components of the laser phase, which represent the effective momentum transfer, are modified by $\pm \chi_k$, and depend on the interferometer branch.
We find from this modification the potential
\begin{equation}
    \hat{V}_{k,\text{G}} = \hbar \frac{h_+}{2} \dot \Lambda_k   k_\text{L} \hat{z} \cos(\omega_\text{G}t + \varphi).
\end{equation}

The combination of both effects is shown by modified light cones due to the altered four wave vector in Fig.~\ref{fig:Potentials}, including the shift of the phase through color code.
Moreover, its effect can be attributed to the Shapiro phase shift~\cite{Badurina2025} in atom interferometers.

As a last building block, we also include the relativistic mass defect.
It attributes to an atom in the excited state a slightly different mass than to one in the ground state, with the mass difference $\Delta m c^2 = \hbar \omega_\text{A}$ being identical to the energy gap due to the mass-energy equivalence.~\cite{Zych2011,Yudin2018,Sonnleitner2018,Schwartz2019,Martinez2022,Asano2024}
As a consequence, any mass-dependent effect experiences a modification, including $\hat{V}_\text{G}$, giving rise to
\begin{equation}
    \hat{V}_{\Delta m,\text{G}} = \Lambda_\omega \frac{h_+}{2}  \frac{\Delta m \omega_\text{G}^2}{2} \hat{z}^2 \cos(\omega_\text{G}t + \varphi).
\end{equation}
\end{subequations}
Moreover, we will later connect this potential to the Einstein phase~\cite{Badurina2025} shift in atom interferometers.
Note that this mass defect in principle has a similar effect on the Newtonian-gravity coupling of a massive particle to Earth's gravity field,~\cite{Sonnleitner2018,Schwartz2019} which has already been the focus of previous studies,~\cite{Ufrecht2020,DiPumpo2021} especially for interferometers generated by single-photon transitions.~\cite{Bott2023}
However, it cancels in the differential measurement schemes considered in this work and can therefore be omitted for protocols without $k$-reversal schemes. 

\section{\label{sec:Principles}Principles of gravitational-wave sensitivity}
In this section, we use the potentials presented above to highlight how atomic quantum sensors acquire their sensitivities to GWs.
To this end, we discuss the key difference between atomic clocks and atom interferometers, which rely on interfering internal states and spatially separated branches, respectively.
We show that the oscillating gravitational potential $\hat{V}_\text{G}$ directly causes a phase shift between branches of matter waves, which is only accessible if the interfering components are in spatial superposition.
Because it is tailored for atom interferometers, this effect is suppressed in clock signals.
However, the laser-phase kick potential $\hat{V}_k$ reads out the position of the atom, giving access to a GW sensitivity of atomic clocks under the condition that it moves on a geodesic.

For our study, we rely on a perturbative phase-computation formalism.~\cite{Ufrecht20202}
It allows for the computation of leading-order effects in phases by integrating over the difference
\begin{equation} \label{eq:perturbationformalism}
    \phi_j = -  \frac{1}{\hbar}\int \dd t \big[V_j^{(1)}(z_1,t)- V_j^{(2)}(z_2,t)\big]
\end{equation}
between perturbative potentials $V_j^{(i)}$ associated with the two interfering components $i=1,2$, which are defined by the coefficients $\Lambda_{k/\omega}^{(i)}$ introduced above.
Here, the interfering component can be abstract, i.\,e., internal, spatial, or a combination of both.
Moreover, the perturbative potentials are evaluated~\cite{Ufrecht20202} at the classical, branch-dependent trajectories $z_i$ generated by the unperturbed part of the respective dynamics.
For now, we focus on $\hat{V}_\text{G}$ and $\hat{V}_k$, and show later and in Appendix~\ref{app.composite_pulses} that these potentials represent the key difference between atomic clocks and atom interferometers.

When we describe atom interferometers, the recoil encoded in $\hat{V}_k$, necessary for generating spatial superpositions, is part of the unperturbed dynamics, while $\hat{V}_\text{G}$ can be treated perturbatively.
In contrast, for atomic clocks the recoil is suppressed in the Lamb--Dicke regime~\cite{Dicke1953,Ludlow2015} through trapping ($k_\text{L} \ll \sqrt{2m\omega_\text{T}/\hbar} $), so that $\hat{V}_k$ can be treated as a perturbation. 
Although in principle, $\hat{V}_\text{G}$ could be considered as a perturbation also for clocks, we incorporate it in this case as part of the unperturbed trajectory, see Appendix~\ref{app.classical_EOMs} for explicit expressions.
The advantage of this approach is that all leading-order effects are attained at first-order perturbation theory given by Eq.~\eqref{eq:perturbationformalism}.

For the effect of $\hat{V}_\text{G}$, we find the phase
\begin{equation}
    \phi_\text{G} = - \frac{h_+}{\hbar}\int \dd t \frac{m \omega_\text{G}^2}{2} \bar{z}(t) \Delta z(t) \cos(\omega_\text{G}t + \varphi),
\end{equation}
defining $z_{1/2}= \bar{z} \pm \Delta z/2$ as the spatial mean and difference of the interfering components of the unperturbed trajectory.
For atom interferometers, the recoil induced by a non-perturbative $\hat{V}_{k}$ leads to a considerable separation $\Delta z$ between the interfering branches, providing access to the phase induced by $\hat{V}_\text{G}$.
On the other hand, for atomic clocks, the recoil is suppressed and $\hat{V}_{k}$ is treated as a perturbation.
Consequently, the interfering components are not spatially separated  ($\Delta z \cong 0$) and as a consequence, no access to $\hat{V}_\text{G}$ is provided for atomic clocks to leading order.

However, $\hat{V}_{k}$ reads out the position of the atom, yielding
\begin{equation}
    \phi_k = \int \dd t \Delta \dot \Lambda_k k_\text{L} \bar z(t),
\end{equation}
where $\Delta \dot \Lambda_k$ is the difference of the delta-pulse sequences defining the interfering components.
This phase is only susceptible to GWs if $\bar z(t)$ is modified by the GW itself, most prominently by a freely falling trapping potential following a geodesic when it interacts with the GW.
If the trap is, however, fixed in a terrestrial setup, which constitutes the realistic case,~\cite{Paul1990,Zhukas2021,Martinez2022} the dependence of $\bar z(t)$ on the GW is suppressed, so that to leading order no GW sensitivity arises.

As a result, the leading-order sensitivity of atom interferometers, both light-pulse and guided ones, originates from the spatial superposition of their branches, while clocks are only sensitive if the atom or the trap center are falling freely, which is in agreement with proposed space-based experiments~\cite{Kolkowitz2016,Norcia2017,Ebisuzaki2020,Fedderke2022} but not realistic on Earth.

\begin{figure}
    \centering
    \includegraphics[width=\columnwidth]{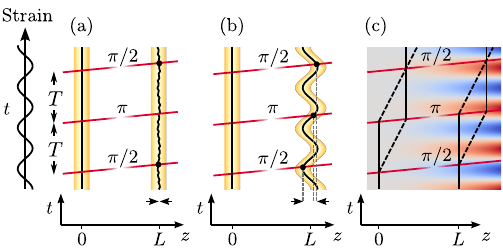}
    \caption{
    Spacetime diagrams of differential setups for detecting a GW with the strain shown on the left.
    The two sensors are separated by a distance $L$ and driven by common $\pi/2$ and $\pi$ pulses (red).
    Traps are drawn in yellow.
    (a) For two stationary traps, the atom in the trap away from the origin oscillates with a suppressed amplitude (shown on the bottom) due to the strong confinement of the trap.
    (b) In contrast, when moving on geodesics, both the atom and trap move as a consequence of the equivalence principle, so that the oscillation amplitude (shown on the bottom) is larger and in phase with the GW.
    (c) Light-pulse atom interferometers, shown as Mach--Zehnder schemes, probe the potential $\hat{V}_\text{G}$ induced by the GW (density plot).
    }
    \label{fig:ClockvsGrad}
\end{figure}

Since such GW signals are always competing with parasitic signals from noise, one has to resort to differential setups~\cite{Mitchell2022,Carlton2025}, which can suppress many noise effects.
To this end, we infer the difference $\delta\phi_j=\phi_{j,a}-\phi_{j,b}$ between two experiments $a$ and $b$, separated by the distance $L= c \tau $, which causes a time delay $\tau$ due to the finite speed of the interrogating light pulses.
Hence, we consider the same pulse sequence acting on both interferometers only delayed by $\tau$.
Figure~\ref{fig:ClockvsGrad} shows a comparison of atomic clocks and atom interferometers.
Our treatment requires that the expansion of the metric in our reference frame, namely, the laboratory, and the associated effective potentials are also valid on the scale of separation of both sensors.

As such, for the kick potential, we find the differential phase
\begin{equation}
    \delta \phi_k = \int \dd t k_\text{L} \big[\Delta \dot \Lambda_{k,a} \bar z_a-\Delta \dot \Lambda_{k,b} \bar z_b\big]. 
\end{equation}
In the following, we assume $\Delta \dot \Lambda_{k,a} (t)= \Delta \dot \Lambda_{k,b}(t-\tau) \cong \Delta \dot \Lambda_{k,b}- \tau \Delta \ddot \Lambda_{k,b}$ due to small $\tau$.
In the remainder of this section, we consider only pulses coming from the left, which means that no large-momentum-transfer or $k$-reversal schemes\cite{McGuirk2002} are possible but are included by a more general treatment in Sec.~\ref{sec:Enhance}.
As a result,
\begin{equation}
    \delta \phi_k \cong \int \dd t \Delta \dot \Lambda_{k,b} k_\text{L} L \left[ \frac{\delta \bar z}{L}+ \frac{\dot{\bar{z}}_a}{c}\right]  \cong \int \dd t \Delta \dot \Lambda_{k,b} k_\text{L}  \delta \bar z,
\end{equation}
where we defined $\delta \bar z =  \bar z_a-\bar z_b$.
In the last step, we assumed $\delta \bar z/L \gg \dot{\bar{z}}_a/c$, valid for the case studied below if $k_\text{G} L\ll 1$.
If this phase is to be used for GW detection, $\delta \bar z$ must be modified by the wave and depend on $h_+$.
For additional physical intuition, we integrate by parts to find $\delta\phi_k \cong -\int \dd t \Delta\Lambda_{k,b} k_\text{L}\delta\dot{\bar{z}}$ and identify $k_\text{L}\delta\dot{\bar{z}} = \omega_\text{L}\delta\dot{\bar{z}}/c = \delta\omega_\text{L}$ as a GW-induced Doppler shift of the laser frequency.
In this picture, the differential phase can be expressed as an integration over the Doppler shift $\delta\phi_k \cong -\int \dd t \Delta\Lambda_{k,b} \delta\omega_\text{L}$ in analogy to previously described schemes.~\cite{Kolkowitz2016}
This Doppler shift also depends on the direction of the driving light field, in contrast to clock phases that arise from $\hat{V}_\omega$.

For the oscillating harmonic potential induced by GWs, we directly find the differential phase
\begin{align}
\begin{split}
   \delta \phi_\text{G} = - \frac{m h_+ \omega_\text{G}^2}{2\hbar}\int \dd t  \big[& \bar{z}_a(t+\tau) \Delta z_a(t+\tau) \cos\Phi_{t+\tau}\\
   &-\bar{z}_b(t) \Delta z_b(t) \cos\Phi_t\big],
   \label{eq:diffphasephiG}
\end{split}
\end{align}
with $\Phi_t = \omega_\text{G} t + \varphi$.
As a next step, we assume that the distance between the arms of the first sensor is just shifted in time compared to the other sensor, i.\,e. $\Delta z_a(t+\tau)=\Delta z_b(t)$, an assumption that we drop when discussing large-momentum-transfer or $k$-reversal schemes.
Furthermore, we expand $\bar{z}_a(t+\tau) \cong \bar z_a(t) + \dot{\bar{z}}_a(t) \tau = L (\bar z_a/L + \dot{\bar z}_a/c)$ and $\cos \Phi_{t+\tau} \cong \cos\Phi_t- k_\text{G} L \sin \Phi_t$.
Thus, the integrand of Eq.~\eqref{eq:diffphasephiG} becomes
\begin{equation}
    \Delta z_b L \left[ \left( \frac{\delta \bar z}{L}+ \frac{\dot{\bar{z}}_a}{c} \right)\cos \Phi_t- k_\text{G} L \left( \frac{\bar z_a}{L} + \frac{\dot{\bar{z}}_a}{c} \right) \sin  \Phi_t  \right].
\end{equation}
Assuming $k_\text{G} L, k_\text{G} \bar z_a,\dot{\bar{z}}_a/c \ll \delta \bar{z}/L\sim 1$, we obtain the dominant term
\begin{equation} 
\label{eq:diffphasephiGseparation}
   \delta \phi_\text{G} \cong - \frac{m h_+ \omega_\text{G}^2}{2\hbar}\int \dd t \Delta z_b  \delta \bar z \cos \Phi_t
\end{equation}
For a targeted frequency range in the hertz regime, this implies $L\ll c/\omega_\text{G}\sim5 \times 10^7\,$m.
Additionally, the velocity of the differential mean position is much lower than the velocity resulting from the spatial difference between two branches $\delta \dot{\bar z} \ll \Delta \dot{ z}_b $, effectively implying that we can treat the differential mean position as almost constant $\delta \bar z\cong \text{const}$ when applying integration by parts twice, which gives rise to
\begin{equation}
   \delta \phi_\text{G} \cong \frac{m h_+ }{2\hbar} \int \dd t \Delta \ddot z_b\delta \bar z \cos \Phi_t.
\end{equation}

In all cases, we prepare a large separation $\delta \bar z \sim L$ between the two sensors, leaving us with two possible measurement strategies for GWs:
(i) We can directly gain access to the potential $\hat{V}_\text{G}\sim h_+$ by maximizing the spatial separation between both branches in one sensor, for example by resorting to large-momentum transfer, both resulting in a large $\Delta \ddot z_b$ between both interfering components.
This strategy is only accessible with atom interferometers, e.\,g., where $\Delta \ddot z_b~\sim k_\text{L}$ for light-pulse schemes.
In contrast, clocks in the Lamb--Dicke regime $k_\text{L} \ll \sqrt{2m\omega_\text{T}/\hbar} $ experience only a suppressed recoil, hence $\delta \phi_\text{G}$ cannot generate a leading-order GW sensitivity for clocks.
(ii) Alternatively, we can use  $\delta \phi_k$ to determine the atom's position through the spatial part of the atom-light interaction.
Since no spatial superposition is generated in trapped clocks, this is the only mechanism for clocks to become sensitive to GWs, if $\delta \bar z \sim h_+ L$, which implies that the clocks need to follow geodesics and fall freely.

We underline these two strategies by a detailed derivation of the differential phase signals for both cases, i.\,e., for atomic clocks and atom interferometers.

\subsection{Atomic clocks}
We describe light pulses, used to drive internal transitions in clocks, by explicitly modeling $\Delta \dot \Lambda_{\omega,b}=\sum_\ell \lambda_{\omega,\ell}\delta(t-t_\ell)$, where $t_\ell$ is the time of the pulse and we define $ \lambda_{\omega,\ell} = \pm 1$ for $\pi/2$ pulses corresponding to excitation and deexcitation of the internal state, as well as $ \lambda_{\omega,\ell} = -2$ for $\pi$ pulses.
From these specifications, we find the dominant phase
\begin{equation}
    \phi_\omega = -  (\omega_\text{L} - \omega_\text{A}) \sum_\ell \lambda_{\omega,\ell}  t_\ell,
\end{equation}
which is the usual clock phase measuring the laboratory time elapsed during the interferometer scheme.
Since we postpone the treatment of large-momentum-transfer or $k$-reversal techniques to section~\ref{sec:Enhance}, we observe $\Delta \dot \Lambda_{k,b}=\sum_\ell \lambda_{k,\ell}\delta(t-t_\ell)$, where $ \lambda_{k,\ell} = \pm 1$ represents again $\pi/2$ pulses for excitation and deexcitation, while $ \lambda_{k,\ell} = \mp 2$ describes the direction of differential momentum transfers for $\pi$ pulses at $t_\ell$.
We obtain for the phase caused by these pulses
\begin{equation}
     \delta \phi_k =  \sum\limits_{\ell}   \lambda_{k,\ell}k_\text{L}  \delta \bar z(t_\ell), 
    \label{eq:DiffPhasePosReadout}
\end{equation}
which results in a readout of the differential mean position between both sensors.
Since this mechanism is the only one for atomic clocks to become sensitive to GWs, we analyze two special cases: a stationary clock and a freely falling clock.

\subsubsection{Stationary traps}
In the first case, the trap is fixed in the laboratory system, e.\,g., on an optical table (see Fig.~\ref{fig:ClockvsGrad}).
For the corresponding unperturbed classical trajectory of the atom inside the trap we find
\begin{equation}
  \frac{\delta \bar z}{L}=1 -\frac{ h_+}{2} \frac{ \omega_\text{G}^2 }{\omega_\text{T}^2-\omega_\text{G}^2 }\left[ \cos \Phi_t + \xi(t) \right],
\end{equation}
where $\xi(t)$ includes the initial conditions $\delta \bar{z}(0) = L$ and $\delta \dot{\bar{z}}(0) = 0$, see Appendix~\ref{app.classical_EOMs} for the explicit form.
Note that $\delta \bar z /L=1$ only leads to a phase offset or vanishes for Ramsey sequences and hence can be neglected in our treatment. 
A typical setup~\cite{Leibfried2003,Ludlow2015} for atomic clocks implies that $\omega_\text{G}\ll\omega_\text{T}$, especially in the Lamb--Dicke regime.
However, a possible exotic setup where the trap is resonant to the GW is not discussed here.
With $\xi(t) = \mathcal{O}(1)$, the amplitude of the oscillation $h_+ (\omega_\text{G}/\omega_\text{T})^2$ becomes suppressed, leading to an overall suppressed sensitivity of stationary clocks to GWs as expected.

Note that in principle also clocks with non-stationary trap centers moving along given non-geodesic worldlines~\cite{Hafele1972a,Hafele1972b,DiPumpo2021} are possible. 
However, their GW sensitivity remains suppressed following the same arguments as above.

\subsubsection{Traps on geodesics}
The second case considers an atomic clock, namely the atom and the trap, falling freely in the GW (see Fig.~\ref{fig:ClockvsGrad}).
Due to the equivalence principle,~\cite{Einstein1907} we obtain the classical trajectory
\begin{equation}
  \delta \bar z /L=1 + h_+ [\cos \Phi_t-\cos\Phi_0  + \omega_\text{G} t \sin \Phi_0]  /2,
\end{equation}
where the last term induces a secular behavior linearly increasing with $t$.
This artifact is caused by our idealized initial conditions, where we effectively model the GW as being suddenly switched on at time $t=0$.
Physically realistic GWs have a finite duration and spectral width, so that one must include its pulse shape and wave-packet character, requiring a multi-chromatic description dependent on their Fourier decomposition.
Such a more realistic treatment would mitigate effects resulting from artificial initial conditions and is consistent with approaches to GW analysis based on power spectral densities, which consider realistic waveforms and their spectral content.~\cite{Jaranowski1998,Moore2015,Romano2017}
While this more detailed modeling is beyond this article's scope, it provides a consistent framework explaining the origin of the secular term.

If we again exclude $k$-reversal schemes, we find $\sum_\ell \lambda_{k,\ell}=0$, yielding the kick phase
\begin{equation}
     \delta \phi_k =  h_+ k_\text{L} L \sum\limits_{\ell}   \lambda_{k,\ell} [ \cos \Phi_{t_\ell} + \omega_\text{G} t_\ell \sin \Phi_0]/2. 
\label{eq:TrappedClockGWD}
\end{equation}
As a first example for a pulse sequence, we consider the well-known Ramsey sequence~\cite{Ramsey1950} defined by $\lambda_{k,1}=1$ and $\lambda_{k,2}=-1$, where no other pulses are applied.
The two $\pi/2$ pulses are separated by the interrogation time $T$.
From this sequence we directly infer the differential phase
\begin{equation}
     \delta \phi_k = h_+ k_\text{L} L [\cos \Phi_0 -\cos \Phi_T- \omega_\text{G} T \sin \Phi_0]/2,
\end{equation}
which resembles previous results~\cite{Norcia2017} that focused on clocks placed on geodesic satellites, except for the secular term, which to our knowledge has not been found before.

Since differential phases still contain the unknown initial phase $\varphi$ of the GW, their fluctuations are the actual quantity of interest.
Although in realistic analyses it is defined by more complicated methods connected to actual statistical evaluations,~\cite{Jaranowski1998,Moore2015,Romano2017} for our purpose, it suffices to calculate $\phi_{j,\text{S}}^2 = \int_0^{2\pi} \dd \varphi \delta \phi_j^2 / (2\pi)$, i.\,e., to average the squared differential phase over the unknown phase.
Again assuming $k_\text{G}L \ll 1$ we obtain for the Ramsey sequence
\begin{equation}
    \phi_{k,\text{S}} = \frac{h_+ k_\text{L} L}{2} \left| \sqrt{1 - \cos \omega_\text{G} T -\omega_\text{G} T \sin \omega_\text{G} T + (\omega_\text{G} T)^2/2  } \right|,
\label{eq:RamSignAmpl}
\end{equation}
which yields a sensitivity to GWs to leading order, where the secular term is, however, still contained.

Another frequently used technique is the echo sequence,~\cite{Gharavipour2017} which represents an analog to a Mach--Zehnder atom interferometer~\cite{Kasevich1991,Giltner1995,Cronin2009} discussed in detail below.
It is defined by $\lambda_{k,1}=1$, $\lambda_{k,2}=-2$, and $\lambda_{k,3}=1$, resulting in the phase $\delta \phi_\text{k} =  h_+ k_\text{L}L \sum_\ell \lambda_{k,\ell} \cos\Phi_{t_\ell}/2$, from where we observe that the secular term has vanished in contrast to the Ramsey sequence from Eq.~\eqref{eq:RamSignAmpl}.
We find for the Echo sequence
\begin{equation}
   \delta \phi_\text{k} = -2 h_+ k_\text{L}L \cos\Phi_T \sin^2 \frac{\omega_\text{G} T}{2},
\end{equation}
resulting in the signal amplitude
\begin{equation}
\label{eq:DiffPhaseEcho}
    \phi_{k,\text{S}} = \sqrt{2}h_+ k_\text{L} L \sin^2 \frac{\omega_\text{G} T}{2},
\end{equation}
susceptible to GWs to leading order, with an even cleaner form compared to the Ramsey sequence.

To give an order of magnitude estimate for this phase, we assume optical transitions with $k_\text{L}=10^7\,$m$^{-1}$, e.\,g., for strontium,~\cite{Hu2017} and a baseline~\cite{Abend2023,Abdalla2025} up to $L=10^3\,$m in terrestrial setups.
On resonant mode with $\omega_\text{G}T=\pi$, where the targeted frequency range is typically in the hertz regime and $T\sim 1 \,$s,
we arrive at $\phi_{k,\text{S}}\sim h_+ 10^{10}\,$rad. 
Unfortunately, due to the miniature gravitational strain $h_+$, such an unenhanced phase measurement is not competitive, which is why we present a more detailed noise analysis in the context of composite interrogation protocols in Sec.~\ref{sec:Enhance}, which enhance the sensitivity.

In conclusion, we found for differential phases the same result as for the single phases before, rendering terrestrial GW detectors based on trapped atomic clocks unrealistic but, indeed, suitable for targeted space missions.~\cite{Kolkowitz2016,Ebisuzaki2020,Fedderke2022}

Note that, in principle, one could try to build an untrapped and freely falling atomic clock,~\cite{Kasevich1989,Wynands2005,DiPumpo2023} which would rely on the position readout from Eq.~\eqref{eq:DiffPhasePosReadout} for GW detection and where the geodesic nature of its differential mean trajectory would arise naturally.
However, since such clocks, which are already used in the microwave regime,~\cite{Kasevich1989,Wynands2005} can only be operated without optical recoils (and thus with negligible spatial superposition), their GW sensitivity would be suppressed as well.

\subsection{Atom interferometers}
In the previous section, we observed that clocks are to leading order only sensitive to GWs if they move on geodesics induced by these waves.
For light-pulse atom interferometers,~\cite{Kasevich1991,Cronin2009} this condition is naturally satisfied.
Moreover, in this case, the spatial part of the atom-light interaction with its non-negligible recoil is generating a spatial superposition, which gives access to $\delta \phi_\text{G}$, i.\,e., a direct measurement of GW potential.
This is also the case if the atom interferometers are not freely falling but guided.
In the following, we analyze both the GW sensitivities of light-pulse atom interferometers and guided atom interferometers and compare their signals with those in atomic clocks.

\subsubsection{Light-pulse atom interferometers}
In contrast to clocks, for light-pulse atom interferometers, $\hat{V}_k$ cannot be treated as a perturbation but must be considered as part of the unperturbed dynamics inducing spatial separations between both branches.
The unperturbed dynamics is dominated by this large spatial separation induced by $m \Delta \ddot z_b/\hbar = \Delta \dot\Lambda_{k,b} k_\text{L}$, while $\hat{V}_\text{G}$ can be treated as a perturbation.
For a pulse sequence $\Delta \dot \Lambda_{k,b}=\sum_\ell \lambda_{k,\ell}\delta(t-t_\ell)$ we obtain
\begin{equation}
   \delta \phi_\text{G} \cong \frac{h_+ k_\text{L} L}{2}\sum_\ell\lambda_{k,\ell} \cos \Phi_{t_\ell},
\end{equation}
indicating that the result for a freely falling trapped clock from Eq.~\eqref{eq:TrappedClockGWD} is similar to the differential phase of light-pulse atom interferometers.
For a Mach--Zehnder sequence, defined by $\lambda_{k,1}=1$, $\lambda_{k,2}=-2$, $\lambda_{k,3}=1$, we find the explicit expression
\begin{equation}
   \delta \phi_\text{G} \cong -2 h_+ k_\text{L} L\sin^2{\frac{\omega_G T}{2}}\cos{\Phi_T},
\end{equation}
with signal amplitude
\begin{equation}
   \phi_\text{G,S} \cong \sqrt{2} h_+ k_\text{L} L\sin^2{\frac{\omega_G T}{2}},
\end{equation}
which exactly reproduces the known result,~\cite{Dimopoulos2009,Graham2013,Badurina2025} identical to the expression of the echo-sequence clock from Eq.~\eqref{eq:DiffPhaseEcho}.

\subsubsection{Guided atom interferometers}
Instead of using freely falling atoms which only interact with electromagnetic fields through short light pulses, it is also possible to build atom interferometers from atoms being subject to trapping potentials, leading to guided atom interferometry.~\cite{Dumke2002}
Such setups often use double well,~\cite{Schumm2005} tractor,~\cite{Duspayev2021} or tweezer~\cite{Nemirovsky2023} techniques to spatially split and recombine the interferometer branches, with holding segments, e.\,g., through wave guides,~\cite{Kovachy2010} between them.
Guided atom interferometers have been demonstrated for inertial sensing,~\cite{Kovachy2010} most prominently for Sagnac rotation sensors,~\cite{Beydler2024} and are at the heart of atomtronics.~\cite{Amico2021}

For two guided atom interferometers,~\cite{Berrada2013,Ke2018,Premawardhana2024} we assume that each of them consists of two initially colocated traps.
After preparation, a splitting mechanism creates a spatial superposition.
The corresponding acceleration of the wave packet on each branch is chosen such that a constant differential mean distance $\delta \bar{z}=L$ between two guided atom interferometers is generated, which implies the same procedure in both atom interferometers. 
We assume the acceleration to be sufficient adiabatic, so that each wave packet follows its respective trap center, leading to a differential phase 
\begin{equation}
   \delta \phi_\text{G} \cong \frac{h_+ m L}{2\hbar} \int \dd t \Delta\ddot{z}_b \cos \Phi_t,
\end{equation}
where no light pulses are necessary.
Our treatment can be generalized to hybrid cases,~\cite{Xu2019,Panda2024} where light pulses are applied in addition to guiding potentials.
However, for purely guided configurations, no light pulses are necessary and only $\hat{V}_\text{G}$ as well as $ \hat{V}_{\Delta m,\text{G}}$ are relevant for the total signal.
Since the latter effect in guided schemes is exactly the same as the former, but with a suppressed prefactor, $\hat{V}_\text{G}$ represents the only relevant potential for such schemes.  

If we compare the differential phase of guided schemes with that of clocks from Eq.~\eqref{eq:DiffPhasePosReadout}, we observe that $h_+$ naturally arises, without using a freely falling trap center.
Hence, in contrast to clocks, guided atom interferometers, although not moving on geodesics, gain their sensitivity solely through the spatial separation of their branches.

As an example, we consider two traps which are initially accelerated by $\pm a/2$ and $\Delta \ddot z_b = a$ for a time $T_a/2$ and immediately afterwards decelerated by $\mp a/2$ and $\Delta \ddot z_b = -a$ for a time $T_a/2$. 
This process is followed by a holding segment for another time $T$, where $\Delta \ddot z_b = 0$, before the traps are moved back to the original height using the sequence opposite as before, \ie{} first accelerating the traps $\Delta \ddot z_b =- a$ over a period $T_a/2$ and afterwards decelerating the traps $\Delta \ddot z_b = a$ over another period $T_a/2$.
Although in this treatment instantaneous changes of the acceleration are applied, we assume sufficiently adiabatic processes, such that the rate of the change of the acceleration is always much smaller than the oscillation period of the trap, $\Delta \dddot z _b /\left(\omega_\text{T}\Delta \ddot z_b\right)\ll1$, suppressing any corrections to this model.
For this setup, we obtain the differential phase
\begin{align}
\begin{split}
   \delta \phi_\text{G} \cong -\frac{4h_+ m a L}{\hbar\omega_\text{G}} \sin^2 \frac{\omega_\text{G}T_a}{4}\sin \frac{\omega_\text{G}\left(T+T_a\right)}{2} \cos \Phi_{T/2+T_a},
\end{split}
\end{align}
which scales inversely to $\omega_\text{G}$, while the transferred momentum corresponds to $\hbar k_a= m a T_a $.
Moreover, we find the signal amplitude 
\begin{align}
\begin{split}
   \phi_\text{G,S} \cong\frac{h_+}{\sqrt{2}} k_a L \left|\sinc{\frac{\omega_\text{G}T_a}{4}}\sin{\frac{\omega_\text{G}T_a}{4}}\sin{ \frac{\omega_\text{G}(T_a+T)}{2}}\right|.
\end{split}
\end{align}
Of course in a realistic treatment, fluctuations of the guiding potential~\cite{Premawardhana2024} and, thus of $a$ have to be included, which is beyond the scope of this work.

\section{\label{sec:Enhance}Enhancement by composite interrogation protocols}
\begin{figure}
    \centering
    \includegraphics[width=\columnwidth]{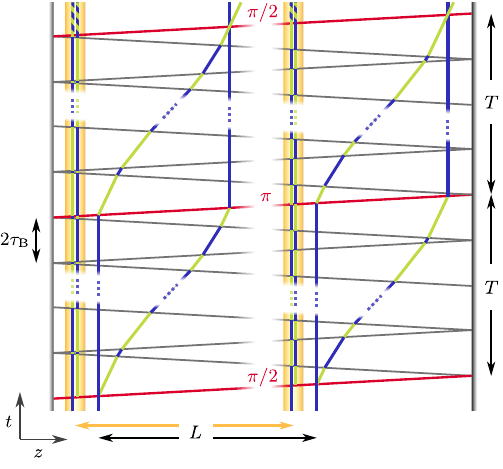}
    \caption{
    Composite interrogation protocol using single-photon transitions for both atom interferometers and atomic clocks (traps highlighted in yellow).
    The large-momentum-transfer Mach--Zehnder interferometer consists of generalized beam-splitter pulses (a $\pi/2$ pulse in red accompanied by $N-1$ gray $\pi$ pulses) and a generalized mirror pulse (a $\pi$ pulse in red sandwiched by $N-1$ gray $\pi$ pulses). 
    The resonance is tuned such that, except for the red pulses, only one arm is addressed and the internal state is changed (blue: ground state and green: excited state).
    The same sequence is used for clocks, however, in that case, both interfering components are always resonant.
    The interrogation time $T$ separates the red pulses, where gray pulses are separated by a time $2\tau_\text{B}$.
    }
    \label{fig:GradSequ}
\end{figure}

The sensitivities of both clocks and atom interferometers can be enhanced by composite interrogation protocols.~\cite{Norcia2017}
For atom interferometers, the enclosed phase-space area can be enhanced by large-momentum transfer,~\cite{Rudolph2020,Gebbe2021,Berg2015} increasing their sensitivities.~\cite{Graham2013,Abend2023,Abdalla2025}
In contrast, hyper-Ramsey or hyper-echo~\cite{Yudin2010,Zanon2015,Hobson2016} inspired schemes~\cite{Zanon2025} for clocks are used to suppress noise by dynamical decoupling~\cite{Viola1999,Suter16} but can lead to a sensitivity enhancement as well, a similarity already observed previously.~\cite{Norcia2017}
The similar mechanism causes an increase in sensitivity through an increasing number of atom-light interaction points.
The effect of this interaction is determined by the internal state prior to it, as well as the phase of the driving light field.
To visualize this common principle, we consider the scheme and pulse sequence presented in Fig.~\ref{fig:GradSequ}.
This scheme constitutes a generalization of both the Mach--Zehnder and echo sequences considered in Sec.~\ref{sec:Principles}.
It explicitly allows for pulses from the opposite direction, e.\,g., necessary for atom interferometers with single-photon transitions but also for atomic clocks.

Large-momentum-transfer sequences of atom interferometers are designed to create large arm separations by transferring a high number of photon recoils, which enhances the sensitivity to GWs via the GW potential.
During such sequences, the laser interacts only with one arm of the spatial superposition.
To successively increase the transferred momentum, the direction of the driving laser must be reversed between subsequent pulses, because photon absorption and stimulated emission impart recoils in opposite directions.  
In contrast, for the clock sequence, there is no Doppler detuning, so the light interacts with both internal states simultaneously.
Two closely spaced $\pi$ pulses would then effectively act as a $2\pi$ pulse, producing no net phase shift.
However, these pulses are phase sensitive: if the direction of the driving light field is reversed between such pulses, the phases at their interaction points add constructively. 
Generalizing this to multiple pulses creates the large-momentum-transfer and hyper-echo protocols that enhance the signal by coherently adding up such phase contributions.
The opposite directions of the pulses highlight the difference between $\Delta \Lambda_\omega$ and $\Delta \Lambda_k$, because now the momentum transfer in different directions becomes most evident, forbidding certain assumptions from Sec.~\ref{sec:Principles}.
We will also explicitly discuss the effects of all other potentials in addition to $\hat{V}_\text{G}$ and $\hat{V}_k$.
For a detailed derivation of all phases, we refer to Appendix~\ref{app.composite_pulses}.

\subsection{Hyper-echo scheme for clocks}
As before, due to the suppressed recoil in clocks, we treat $\hat{V}_k$ as a perturbation when computing their phase signals.
Moreover, we again use $\hat{V}_\text{G}$ to infer the unperturbed part of their classical dynamics.
As a result, we find the differential phase 
\begin{equation}
    \delta \phi_k= \frac{h_+}{2} k_\text{L}L \sum\limits_\ell \lambda_{k,\ell} [\cos\Phi_{t_\ell^a} + \omega_\text{G} t_\ell^a \sin \Phi_0]
\end{equation}
originating from $\hat{V}_k$ for two spatially separated clocks, with $\bar z_b \ll \bar z_a = L + z_b \cong L$ and where $t^a_\ell$ is the time of the interaction with the clock far from the origin of the laboratory frame.
We considered clocks and traps that follow geodesics according to Appendix~\ref{app.classical_EOMs}, as we will throughout the remainder of this subsection, since this configuration yields a sensitivity to GWs.
For a hyper-echo sequence, we have a symmetric pulse sequence such that $\sum_\ell \lambda_{k,\ell} t_\ell^a =0$ and find
\begin{equation}
\label{eq:DiffPhasekHyper}
    \delta \phi_k=\frac{h_+}{2} k_\text{L}L \sum\limits_\ell \lambda_{k,\ell} \cos\Phi_{t_\ell^a} 
\end{equation}
for the differential phase from $\hat{V}_k$. 
Relying on Appendix~\ref{app.composite_pulses}, we find under the same conditions that the mass defect yields the differential phase
\begin{equation}
\label{eq:DiffPhaseMassHyper}
    \delta\phi_{\Delta m, \text{G}} = \frac{1}{4} h_{+} k_\text{A}  k_\text{G} L^2  \sum_\ell \lambda_{\omega,\ell} \sin \Phi_{t_\ell^a},
\end{equation}
while we obtain from the modified laser kick
\begin{equation}
\label{eq:DiffPhaseFreqHyper}
    \delta \phi_{k, \text{G}} = -\frac{h_{+}}{2} k_\text{L} L \sum\limits_\ell \lambda_{k, \ell} \cos \Phi_{ t_\ell^a } = - \delta \phi_k,
\end{equation}
and from the modified laser frequency
\begin{align}
\begin{split}
\label{eq:DiffPhaseWaveHyper}
    \delta \phi_{\omega, \text{G}} =& \frac{h_+}{2} \frac{k_\text{L}}{k_\text{G}} \sum_\ell  \lambda_{\omega, \ell} [  \sin \Phi_{t_\ell^a}- \sin \Phi_{t_\ell^b} ] - \frac{k_\text{L}}{k_\text{A}}\delta\phi_{\Delta m, \text{G}}.
\end{split}
\end{align}

In the resonant case $k_\text{A}\cong k_\text{L}$, and under the assumption $\sin \Phi_{t_\ell^a}- \sin \Phi_{t_\ell^b}\cong k_\text{G}L\cos\Phi_{t_\ell^b} $, we find that, without the additional hyper-echo pulses (i.\,e., $\lambda_{\omega,\ell}=\lambda_{k,\ell}$), $\delta \phi_{\omega, \text{G}} \cong -\delta \phi_{k, \text{G}} - \delta\phi_{\Delta m, \text{G}}$.
Therefore, $\delta \phi_{\omega, \text{G}}+\delta \phi_{k, \text{G}} + \delta\phi_{\Delta m, \text{G}} \cong 0$, which justifies omitting the laser phases in Sec.~\ref{sec:Principles}.

Note that for single-photon transitions, the difference between $k_\text{A}$ and $ k_\text{L}$ is a relativistic correction~\cite{Bott2023} and is of higher order, which we omit in this work.
Similarly, finite speed of light on the scale of a single atomic sensor induces phases of similar order,~\cite{Tan2016,DiPumpo2023,Niehof2025} and for clocks it is further suppressed due to the absences of a spatial superposition.
As a result, we assume $k_\text{A}\cong k_\text{L}$ in the following. 
For the total differential phase $\delta \phi_k+\delta \phi_{k, \text{G}}+\delta \phi_{\omega, \text{G}} +\delta\phi_{\Delta m, \text{G}}$ in a hyper-echo clock scheme, we find
\begin{equation}
    \delta \phi = \frac{h_+}{2} \frac{k_\text{L}}{k_\text{G}} \sum_\ell  \lambda_{\omega, \ell} [  \sin \Phi_{t_\ell^a}- \sin \Phi_{t_\ell^b} ]. 
\end{equation}
The explicit form of the pulse sequence from Fig.~\ref{fig:GradSequ} uses $N$ pulses for the beam splitters and $2N-1$ pulses for the mirror, which are formally specified in Appendix~\ref{app.composite_pulses}.
This choice leads to
\begin{widetext}
    \begin{align}
\begin{split}
\label{eq:DiffPhaseHyper}
    \delta \phi &=-4 h_{+}N k_\text{L}L \sin  \frac{\omega_\text{G}T}{2} \sinc  \frac{N k_\text{G}L}{2}  \sin  \frac{\omega_\text{G}T - (N-1)k_\text{G}L}{2}     \cos \Phi_{T+\tau/2} + 2 h_{+} k_\text{L} L \sin^2  \frac{\omega_\text{G}T}{2} \sinc \frac{k_\text{G}L}{2} \cos\Phi_{T+\tau/2}.
\end{split}
\end{align}
\end{widetext}

For a small GW vector $N k_\text{G} L \ll 1$, we obtain the signal amplitude 
\begin{equation}
    \phi_\text{S} = \sqrt{2}h_{+} (2N-1) k_\text{L} L \sin^2 (\omega_\text{G}T/2), 
\label{eq:HyperSignAmp}
\end{equation}
which includes the original echo-scheme pulses from Eq.~\eqref{eq:DiffPhaseEcho} but is enhanced by a factor $2N-1$, similar to previous works that used a different pulse sequence.~\cite{Norcia2017}

In the limit of many hyper pulses $N \gg 1$, which corresponds to effectively neglecting the contributions from the three original echo pulses, we find 
\begin{align}
\begin{split}
    \phi_\text{S} = &4h_{+} N k_\text{L} L \sinc  \frac{N k_\text{G}L}{2}  \sin \frac{\omega_\text{G}T}{2}  \sin \frac{ \omega_\text{G}T - (N-1) k_\text{G}L}{2}. 
\end{split}
\end{align}
Applying the condition $N k_\text{G} L \ll 1$ from before, we obtain 
\begin{align}
\begin{split}
\label{eq:HyperSignAmpFinal}
    \phi_\text{S} = &2\sqrt{2}h_{+} N k_\text{L} L \sin^2 ( \omega_\text{G}T/2), 
\end{split}
\end{align}
which is the same as the first term in Eq.~\eqref{eq:HyperSignAmp}, isolating the enhancing character of the hyper-echo sequence.

For previously discussed parameters~\cite{Badurina2020,Abe2021,Abend2023,Abdalla2025} of $N=10^4$ pulses and terrestrial separations of up to $L=10^3\,$m, together with an optical wave number of $k_\text{L}=10^7\,$m$^{-1}$, e.\,g. for strontium,~\cite{Hu2017} we obtain from Gaussian error propagation~\cite{DiPumpo2023b,Schach2025} a sensitivity to the strain of $\Delta h_+\sim\Delta\phi_\text{S}/(N k_\text{L}L)\sim10^{-19}\,\text{Hz}^{-1/2}$ when operated in resonant mode with $\omega_\text{G}T=\pi$ and assuming shot-noise limitation with $\Delta\phi_\text{S}=10^{-5}\,\text{rad}\,\text{Hz}^{-1/2}$ phase noise.~\cite{Badurina2020}
Typical frequencies for resonant mode lie in the range of $10^{-2}\,\text{Hz}-10$\,Hz, closing the gap between next-generation optical space-based~\cite{danzmann1996,bayle2022} and optical terrestrial setups.~\cite{Punturo2010}

\subsection{Large-momentum transfer for atom interferometers}
For atom interferometers, we again consider $\hat{V}_k$ contributing to the unperturbed part due to momentum recoils, while $\hat{V}_\text{G}$ is a perturbation.
Moreover, we assume $L \gg \bar{z}_a-L, \bar{z}_b, \Delta z_a, \Delta z_b$ and neglect higher-order terms in these quantities.
For the time-dependent harmonic potential discussed in Sec.~\ref{sec:Principles}, we obtain after double integration by parts the result
\begin{equation}
    \delta \phi_\text{G} = \frac{m h_{+}L}{2\hbar} \int \! \dd t \, \Delta \ddot z_a \cos \Phi_t.
\end{equation}

So far, this expression is completely general.
However, since for guided atom interferometers there are no distinguished large-momentum-transfer schemes, we focus in the following on light-pulse configurations.
In this case, we can use $m \Delta \ddot z_a /\hbar = \Delta \dot \Lambda_{k, a} k_\text{L}$ as well as a general pulse sequence $\Delta \dot \Lambda_{k, a} = \sum_\ell \lambda_{k, \ell} \delta (t-t_\ell^a)$, obtaining
\begin{equation}
    \delta \phi_\text{G}=\frac{h_+}{2} k_\text{L}L \sum\limits_\ell \lambda_{k,\ell} \cos\Phi_{t_\ell^a}, 
\end{equation}
which is the same as the result from $\hat{V}_k$ for clocks from Eq.~\eqref{eq:DiffPhasekHyper} with centers moving on geodesics.
In addition, for this configuration, the differential phases resulting from all other potentials are exactly the same as for clocks and are listed in Eqs.~\eqref{eq:DiffPhaseMassHyper} -- \eqref{eq:DiffPhaseWaveHyper} (see Appendix~\ref{app.composite_pulses}).

Thus, we showed that the origin of the difference in the GW sensitivity between atomic clocks and atom interferometers, both guided and light-pulse ones, is indeed given by $\hat{V}_k$ and $\hat{V}_\text{G}$.
As such, we find for the total differential phase of a large-momentum-transfer Mach--Zehnder atom interferometer
\begin{align}
\begin{split}
    \delta \phi =& -2 h_{+}N k_\text{L}L \sin  \frac{\omega_\text{G}T}{2} \sinc  \frac{N k_\text{G}L}{2}\\
    &\times\sin  \frac{\omega_\text{G}T - (N-1)k_\text{G}L}{2}     \cos \Phi_{T+\tau/2},
\end{split}
\end{align}
which has also been derived with different methods before.~\cite{Badurina2025}
It is similar to the first term of the result from hyper-echo clocks in Eq.~\eqref{eq:DiffPhaseHyper}, albeit not identical.
For $N k_\text{G} L \ll 1$, we get the familiar signal amplitude 
\begin{equation}
\label{eq:FinalResMZIAI}
    \phi_\text{S} = \sqrt{2}h_{+} N k_\text{L} L \sin^2 (\omega_\text{G}T/2), 
\end{equation}
which represents almost half of the approximated signal for hyper-echo clocks from Eq.~\eqref{eq:HyperSignAmp}.
Finally, for the large-pulse-number limit $N \gg 1$, we find, apart from a factor of $2$, the same result as for clocks in the same limit as in Eq.~\eqref{eq:HyperSignAmpFinal}.
As a consequence, the sensitivity to the strain $h_+$ of large-momentum-transfer atomic GW sensors lies in the same order of magnitude of $\Delta h_+\sim\Delta\phi_\text{S}/(N k_\text{L}L)\sim10^{-19}\,\text{Hz}^{-1/2}$ as the one of clocks on geodesics, if the same parameters for pulses $N=10^4$, separation $L=10^3\,$m, wave number $k_\text{L}=10^7\,$m$^{-1}$, phase noise~\cite{Badurina2020} $\Delta\phi_\text{S}=10^{-5}\,\text{rad}\,\text{Hz}^{-1/2}$, as well as resonant mode in the targeted frequency range are used.

\section{Conclusions}
In this study, we analyzed and compared the GW sensitivities of atom interferometers and atomic clocks in the laboratory frame.
For this analysis, we took all leading-order potentials from different GW-induced origins into account, including a direct coupling to massive particles, the mass defect, and couplings to the frequency of light and to its wave vector.

We observed that the key ingredient for GW sensitivity of atom interferometers, both light-pulse and guided ones, is their spatial superposition, enabling access to the direct matter coupling to GWs through a time-dependent harmonic potential.
This coupling mechanism is not accessible to clocks, since they inherently suppress spatial superposition, either by trapping in the Lamb--Dicke regime or by recoilless pulses.
However, we showed that light pulses, also necessary for generating internal superpositions, read out the position of the atom, using the wave vector of the pulse as the relevant length scale.
Although this effect still offers no access to GWs with free clocks using recoilless pulses, it opens the possibility to imprint geodesic motion in trapped clocks with large effective wave vectors.
We found that this mechanism requires traps and atoms to move on geodesics of the GW, which is a natural situation for satellite free-fliers but unrealistic in terrestrial setups.

Moreover, we derived composite interrogation protocols with multiple points of atom-light interaction for sensitivity enhancement, which are also discussed~\cite{Arvanitaki2018,Badurina2023,DiPumpo2023b} in the context of dark matter.~\cite{Derr2023}
We showed that generalizations of both Ramsey or spin-echo schemes for clocks as well as Mach--Zehnder schemes for atom interferometers can be expressed on a common ground.
To this end, we transferred exactly the same pulse sequence of the Mach--Zehnder scheme to clocks, which has an additional benefit because echo-type configurations suppress DC noise, distinguishing our results from previously introduced pulse sequences.~\cite{Norcia2017} 

We related individual contributions of the differential phase in atom interferometers to previous results~\cite{Badurina2025} that identified Doppler, Shapiro, and Einstein terms, which also have analogs in optical GW detectors.~\cite{Lee2025}
In addition, we showed similarities and differences to the differential phases of clocks.
These results could be generalized to multiloop geometries~\cite{Graham2016b,DiPumpo2023b,Schach2025} for resonant-mode GW detection, where a comparison between hyper-echo clocks and multiloop atom interferometers rather than large-momentum-transfer ones might be of interest.
While we derived the signal amplitude of the differential-phase fluctuations, our phase expressions can also be connected, by a Fourier transformation into the frequency space, to GW-sensitivity curves based on power spectral densities.~\cite{Jaranowski1998,Moore2015,Romano2017}
Our results can be used for future design choices taking into account different potentials and distinguishing the GW origins of the sensitivities of different atomic quantum sensors.

\section*{Dedication}
We proudly dedicate this article to Ernst Maria Rasel to celebrate his 60th birthday.
His numerous contributions to atom-based quantum sensors, particularly for fundamental-physics and matter-wave optics applications, have laid the foundation for our work.
Wishing you all the best, Ernst!

\begin{acknowledgments}
The authors are grateful to W. P. Schleich for his stimulating input and continuing support.
The authors also thank P. Schach and A. Friedrich as well as the QUANTUS team for fruitful and interesting discussions.
The QUANTUS project is supported by the German Space Agency at the German Aerospace Center (Deutsche Raumfahrtagentur im Deutschen Zentrum f\"ur Luft- und Raumfahrt, DLR) with funds provided by the Federal Ministry for Economic Affairs and Energy (Bundesministerium f\"ur Wirtschaft und Energie, BMWE) due to an enactment of the German Bundestag under Grant No. 50WM2450D and No. 50WM2450E (QUANTUS-VI).
The authors acknowledge contributions in the form of discussions from the Terrestrial Very-Long-Baseline Atom Interferometry (TVLBAI) proto-collaboration.
S.S. gratefully acknowledges the financial support from the German Academic Scholarship Foundation (Studienstiftung des deutschen Volkes) and the Thomas Weiland Foundation.
F.D.P. is grateful to the financial support program for early career researchers of the Graduate \& Professional Training Center at Ulm University and for its funding of the project ``Long-Baseline-Atominterferometer Gravity and Standard-Model Extensions tests'' (LArGE).
\end{acknowledgments}

\section*{Author declarations}
\subsection*{Conflict of interest}

\noindent The authors have no conflicts to disclose.

\subsection*{Author contributions}
\noindent{\sffamily\small\textbf{Simon Schaffrath}} Conceptualization (supporting); Formal analysis (lead); Methodology (equal); Validation (lead); Visualization (lead); Writing - original draft (supporting); Writing – review and editing (supporting).
{\sffamily\small\textbf{Daniel Störk}} Conceptualization (supporting); Formal analysis (supporting); Methodology (supporting); Validation (supporting); Visualization (supporting); Writing - original draft (supporting); Writing – review and editing (supporting).
{\sffamily\small\textbf{Fabio Di Pumpo}} Conceptualization (equal); Formal analysis (supporting); Methodology (equal); Validation (equal); Visualization (supporting); Writing - original draft (lead); Writing – review and editing (equal).
{\sffamily\small\textbf{Enno Giese}} Conceptualization (equal); Formal analysis (supporting); Methodology (equal); Validation (equal); Visualization (supporting); Writing - original draft (supporting); Writing – review and editing (equal).

\section*{Data availability}
The data that support the findings of this study are available within the article.

\appendix

\section{Effective gravitational-wave potentials}
\label{app.GW-potential}
Here, we derive the quantum-mechanical potentials for atoms based on the coupling of massive particles and light fields to GWs in the laboratory frame.
We omit Earth's gravity and other non-inertial effects to focus on GWs.
To leading order, they behave additive and could be included readily.

\subsection{Metric for gravitational waves in the laboratory system}
In the weak-field approximation $||h_{\mu\nu}|| \ll ||\eta_{\mu\nu}||$, the metric reads $g_{\mu\nu} = \eta_{\mu\nu} + h_{\mu\nu}$, where $\eta_{\mu\nu}$ is the Minkowski metric with sign convention (+ - - -). 
By omitting Earth's linear gravity and rotations, we can use Fermi(-Walker) coordinates~\cite{Fermi1922,Manasse1963,Marzlin1994} spanned along any geodesic to represent our proper reference or laboratory frame.
Thus, a general metric can be written as
\begin{equation} \label{eq:fermimetric}
    \begin{aligned}
        h_{00}(t, \vect{x}) &= - R_{0a0b}(t) x^{a} x^{b} \\
        h_{0k}(t, \vect{x}) &= \frac{2}{3} R_{0abk}(t) x^{a} x^{b} \\
        h_{jk}(t, \vect{x}) &= \frac{1}{3} R_{jabk}(t) x^{a} x^{b}
    \end{aligned}
\end{equation}
to second order in spatial proper coordinates $x^a$.
Here, $x^{0}/c = t$ is the laboratory time of the proper reference frame and $\vect{x}$ are the spatial coordinates corresponding to positions in the laboratory.
As such, Greek indices include $\left(0,x,y,z\right)$, while Latin ones cover $\left(x,y,z\right)$.
Moreover, $R_{\mu\nu\rho\sigma}(t) = \frac{1}{2}\left( \partial_\rho \partial_\nu h_{\mu \sigma} + \partial_\mu \partial_\sigma h_{\nu \rho} - \partial_\sigma \partial_\nu h_{\mu \rho} - \partial_\rho \partial_\mu h_{\nu \sigma} \right)$ is the Riemann curvature tensor evaluated on the worldline of the laboratory at time $t$. 

We start from a metric induced by GWs in transverse-traceless (TT) gauge.~\cite{Misner1973}
If we consider a GW, generated by a distant source,~\cite{Maggiore2000,Sathyaprakash2009,abbott2016,Caprini2018} which propagates in ${x}^\prime$ direction with frequency $\omega_\text{G} =c k_\text{G}$ and initial phase $\varphi$, the metric perturbation
\begin{equation} \label{eq:ttmetricgw}
    \left(h_{{\mu}^\prime{\nu}^\prime}\right) = 
    \begin{pmatrix}
    0 & 0 & 0 & 0 \\
    0 & 0 & 0 & 0 \\
    0 & 0 & h_{+} & h_{\times} \\
    0 & 0 & h_{\times} & -h_{+}
    \end{pmatrix} \cos{\left(\omega_\text{G} t^\prime - k_\text{G} {x}^\prime + \varphi\right)},
\end{equation}
depends on the strains $h_{+}$ and $h_{\times}$ corresponding to different polarizations.
Here, indices with a prime denote quantities in the TT frame.
With this metric, we calculate the corresponding components of the Riemann tensor $R_{{\mu}^\prime{\nu}^\prime{\rho}^\prime{\sigma}^\prime}$. 
Because in the weak-field approximation the components of the Riemann tensor are invariant~\cite{Carroll2004,Maggiore2007} under small coordinate transformations, we use the identity $R_{{\mu}^\prime{\nu}^\prime{\rho}^\prime{\sigma}^\prime}=R_{\mu\nu\rho\sigma}$.
In our study, we consider only interferometers operating orthogonally to the direction of the propagation of the GW, such that we need only the metric in laboratory coordinates at $x=y=0$.
Using Eq.~\eqref{eq:fermimetric} and neglecting Earth's gravity gradients, we calculate the metric in Fermi coordinates and find
\begin{equation} \label{eq:fermimetricgw}
    \left( h_{\mu\nu} \right) = \begin{pmatrix}
    3 & -2 & 0 & 0 \\
    -2 & 1 & 0 & 0 \\
    0 & 0 & 0 & 0 \\
    0 & 0 & 0 & 0
    \end{pmatrix} \frac{h_{+}}{6} \left(k_\text{G} z\right)^2 \cos{\left( \omega_\text{G} t + \varphi \right)},
\end{equation}
where we chose $\left(t^\prime=t,\vect{x}^\prime=0\right)$ as spacetime point of origin of the laboratory frame.
In principle, the proper reference frame is expanded at the proper time of the observer in the original TT frame.
However, first, possible corrections from the proper time are beyond the order of approximation in this work.
Second, by neglecting Newtonian gravity, for a static observer in the TT frame the proper time corresponds exactly to the coordinate time $t^\prime$.
Moreover, a constant spatial origin different from $\vect{x}^\prime=0$ could always be absorbed into $\varphi$ and is therefore irrelevant for our purposes.

Note that if Earth's linear gravity and rotations should be included, we need to expand the laboratory frame along an Earth-suspended worldline instead of a geodesic one.
However, GWs yield to leading order no contributions to the four-acceleration arising from the expansion along such a worldline,~\cite{Maluf2008,Kajari2009} which normally gives rise to the linear acceleration and trivial rotations.
As a consequence, including them would only lead to additive effects, which are neglected in our study similar to Earth's gravity gradients.

\subsection{Potentials acting on center of mass of massive particles}
Next, we focus on the coupling to massive particles.
Because we consider atoms composed of massive constituents,~\cite{Parker1980,Lammerzahl1995,Pachucki2007,Perche2021,Asano2024} each constituent couples to the GW through the metric.
To describe atoms, we change from the coordinates of the individual constituents to center-of-mass (c.m.) and relative coordinates.~\cite{Close1970,Krajcik1974}
The c.m. motion of the particle is to leading order determined~\cite{Lammerzahl1995,Schwartz2019,Janson2025} by the analog of the Newtonian potential in a GW and, therefore, the rest mass couples to the $h_{00}$ component of the metric, so that the effective potential is $mc^2 h_{00}/2$. 
Moreover, relativistic cross-couplings between c.m. and relative coordinates imply the mass defect,~\cite{Zych2011,Sonnleitner2018,Loriani2019,Schwartz2019,DiPumpo2021,Janson2025} where due to the mass-energy equivalence atoms in different internal states will have a different mass.
For two internal states of energy difference $\Delta m c^2$, this effect leads to the additional potential $\Lambda_\omega \Delta mc^2 h_{00}/2$, where $\Lambda_\omega=\pm 1/2$ if the atom is in the excited and ground state, respectively.
In addition, the GW coupling to the pure relative motion generates corrections to the internal energy levels of atom,~\cite{Wanwieng2023,Chen2024,Wanwieng2025} but due to small relative distances and the atomic configurations, these effects are strongly suppressed and omitted in this work.

As a result, for the metric from Eq.~\eqref{eq:fermimetricgw} with $\omega_\text{G}=c k_\text{G}$ at $x=y=0$, we obtain for the c.m. and mass defect the effective potentials
\begin{align}
    \hat{V}_\text{G} = \frac{h_+}{2} \frac{m \omega_\text{G}^2}{2} \hat{z}^2 \cos(\omega_\text{G}t + \varphi)
\end{align}
and
\begin{align}
    \hat{V}_{\Delta m,\text{G}} =\Lambda_\omega  \frac{h_+}{2} \frac{\Delta m \omega_\text{G}^2}{2} \hat{z}^2 \cos(\omega_\text{G}t + \varphi),
\end{align}
where $\hat{z}$ is the quantized c.m. position operator.

\subsection{Light propagation in gravitational waves}
Apart from massive particles, also gauge fields like electromagnetic waves couple to gravity.~\cite{Ruggiero2025}
Since we consider configurations where the atom is at a distance from the laser sources and the laser's wavelength is small compared to the curvature of the metric, we can resort to a geometrical-optics ansatz.~\cite{Misner1973,Mieling2021}
In this approximation, effects from polarizations and amplitudes are suppressed compared to those from the light's phase, so that the dominant gravitational influence on the electromagnetic field is described by the Eikonal equation~\cite{Dolan2018}
\begin{equation}
    g^{\mu\nu} (\partial_\mu \chi)(\partial_\nu \chi) = 0,
\end{equation}
representing the propagation of a light beam with phase $\chi$.

In weak-field approximation, we perform a separation $\chi \cong \chi^{(0)} + \chi^{(1)}$ between the unperturbed phase $\chi^{(0)}$ and a perturbation $\chi^{(1)}$.
Together with $\partial_\mu \chi \cong k_{\text{L,}\mu} + \partial_\mu \chi^{(1)}$ and the unperturbed condition $\eta^{\mu\nu} k_{\text{L,}\mu} k_{\text{L,}\nu} = 0$, we find
\begin{equation} \label{eq:eikonalpert}
    k_\text{L}^\mu \partial_\mu \chi^{(1)} = \frac{1}{2} h_{\mu\nu}k_\text{L}^\mu k_\text{L}^\nu.
\end{equation}
Thus, the light's propagation in positive or negative $z$ direction in Fermi coordinates from above, i.\,e., $( k_{\text{L,}\mu} ) = (k_\text{L}, 0, 0, \mp k_\text{L})$ at $x=y=0$, leads to a total electromagnetic phase
\begin{equation}
    \chi_{\pm}(z, t) = \omega_\text{L} t \mp k_\text{L} z + \phi_0 + \chi_{\pm}^{(1)}(z, t).
\end{equation}
with the unperturbed laser frequency $\omega_\text{L} = c k_\text{L}$ and the initial phase $\phi_0$.
Using this ansatz, the perturbation is determined by
\begin{equation} \label{eq:eikonalpertspec}
    \left( \frac{1}{c} \partial_t \pm \partial_z \right) \chi_{\pm}^{(1)} = \frac{1}{4} h_{+} k_\text{L} (k_\text{G} z)^2 \cos{(\omega_\text{G} t + \varphi)}.
\end{equation}
The particular solution
\begin{equation}
\label{eq:perturb_laser_phase}
    \chi_{\pm}^{(1)} = \chi_\omega \pm \chi_k 
\end{equation}
consists of two contributions
\begin{align}
\chi_\omega = \frac{1}{2} h_{+} \frac{k_\text{L}}{k_\text{G}} \left( \frac{\left(k_\text{G} z\right)^2}{2}  - 1 \right) \sin(\omega_\text{G} t + \varphi)
\end{align}
and
\begin{align}
\chi_k = \frac{1}{2} h_{+} k_\text{L} z \cos(\omega_\text{G} t + \varphi). 
\end{align}
Here, $\chi_\omega$ is independent of the direction of propagation, while the sign in front of $\chi_k$ changes for propagation in the opposite direction in Eq.~\eqref{eq:perturb_laser_phase}.
Because of this direction dependence, we identify $\chi_\omega$ with a frequency modification and $\chi_k$ with a modification of the wave vector.
To fulfill the boundary conditions for the phase, e.\,g., at the laser source or after a reflection at a mirror necessary for composite pulses, we have to add a suitable homogeneous solution, which cancels in differential measurements and is therefore omitted.

After a Power--Zienau--Woolley and a dipole approximation are performed~\cite{Sonnleitner2018,Schwartz2019,Woolley2020,Asano2024}, the electromagnetic field is evaluated at the c.m. position of the atom to leading order.
Upon photon absorption, the phase $\chi$ of the electric field is imprinted on the atom and by that the modification $\chi^{(1)}$ caused by the GW as well.
Hence, the instantaneous interaction of the atom with the laser gives rise to potentials that depend on the c.m. position operator $\hat{z}$, and we find from $\chi_\omega$ the potential
\begin{equation}
    \hat{V}_{\omega,\text{G}} = \hbar \frac{h_+}{2} \dot \Lambda_\omega   \frac{k_\text{L}}{k_\text{G}} \left( \frac{(k_\text{G} \hat{z})^2}{2}-1\right) \sin(\omega_\text{G}t + \varphi).
\end{equation}
Here, we define $\Lambda_\omega = \pm 1/2$ if the atom is in the excited or ground state.
For instantaneous pulses, $\dot \Lambda_\omega$ is a directed series of delta functions at times of the pulses.
From $\chi_k$ we obtain the potential
\begin{equation}
    \hat{V}_{k,\text{G}} = \hbar \frac{h_+}{2} \dot \Lambda_k   k_\text{L} \hat{z} \cos(\omega_\text{G}t + \varphi),
\end{equation}
where $\dot \Lambda_k$ is a series of directed laser pulses as before, but the sign now depends on the direction of the momentum transfer ($+$ up, $-$ down), e.\,g., accounting for branch-dependent large momentum transfer.

\section{Classical equations of motion for clocks}
\label{app.classical_EOMs}
To describe optical clocks, we assume that they are trapped in a harmonic potential with frequency $\omega_\text{T}$ and a classical trap center $z_T$.
Despite the presence of this potential, the whole setup may be subject to accelerations, including a possible movement~\cite{Hafele1972a,Hafele1972b,DiPumpo2021} of the laser system generating the trap.
For this movement, we study two different situations: a stationary trap fixed in the laboratory system and a trap moving on a geodesic generated by a GW.
For our perturbative formalism and for closed interferometer geometries, classical trajectories are sufficient~\cite{Ufrecht20202} for the description of leading-order phase contributions.
To this end, we assume that both interfering components of each clock follow the same unperturbed trajectory, specified below.

\subsection{Trajectory of atoms in stationary traps}
For clocks with a stationary trap center, we define $z_T = \text{const.}$, where we omit Newtonian gravity since its effect on the classical trajectory is common for two clocks in differential setups.
Then, the relevant classical equations of motion, neglecting $\mathcal{O}\left(h_{+}^2\right)$, are given by
\begin{equation}
    \ddot{z} + \omega_\text{T}^2 \left( z-z_T \right) = -\frac{1}{2}h_{+} \omega_\text{G}^2 z_T \cos \Phi_t, 
\end{equation}
which is a differential equation for a driven harmonic oscillator, where the driving force stems from the GW.
Solving this equation with initial conditions $z(t=0) = z_T=z_0$ and $\dot{z}(t=0) = 0$ yields
\begin{align}
\begin{split}
    \frac{z(t)}{z_0} &= 1 - \frac{h_{+}}{2} \frac{\omega_\text{G}^2}{\omega_\text{T}^2 - \omega_\text{G}^2} \left[ \cos \Phi_t + \xi(t) \right], 
\end{split}
\end{align}
where we introduced the abbreviation
\begin{equation}
    \xi = - \frac{ \sqrt{C^2+S^2 }}{\omega_\text{T}}
     \cos \left[ \omega_\text{T}t + \arctan \left( C, S \right) \right],
\end{equation}
with 
$C=\omega_\text{T} \cos \varphi$ and 
$S=\omega_\text{G} \sin \varphi$.
From this result, we can directly anticipate the suppression of the GW signal discussed in Sec.~\ref{sec:Principles}.

\subsection{Trajectory of atoms in geodesic traps}
Next, we focus on clocks where the trap center follows a geodesic induced by a GW.
We find the geodesic equation 
\begin{equation}
    \ddot{x}_T^\mu(t) + \Gamma^{\mu}_{\;\alpha \beta}\left[x_T^\rho(t)\right] \dot{x}_T^\alpha(t) \dot{x}_T^\beta(t) = 0
\end{equation}
with Christoffel symbols $\Gamma^{\mu}_{\;\alpha \beta}$.  
At first, the time derivative is defined with respect to the proper time of the trap.
However, for small velocities and weak fields, i.\,e., in the non-relativistic limit, this derivative can be taken with respect to laboratory time $t$.
Moreover, we find $\dot{x}_T^0 \approx c$ for non-relativistic velocities.
In the context of the weak-field approximation, we separate the trajectory $x_T^\mu=x_\text{(0)}^\mu+x_{(1)}^\mu$ into an unperturbed part $x_{(0)}^\mu(t) = x^\mu_0 + v^\mu (t-t_0)$ and a perturbation $x_{(1)}^\mu (t)$ due to the GW.
By expanding perturbatively, we find
\begin{equation}
    \ddot{x}_{(1)}^\mu + \Gamma^{\mu}_{\;\alpha \beta}\left[x_{(0)}^\rho\right] \dot{x}_{(0)}^\alpha \dot{x}_{(0)}^\beta = 0,
\end{equation}
where we neglect terms that are of higher than linear order in the perturbation.
Integrating this equation and using the initial conditions $ x_T^\mu (t_0) = x^\mu_0$ and $ \dot{x}_T^\mu(t_0) =  v^\mu$ yields
\begin{align}
\begin{split}
     x_{(1)}^\mu (t) = &-\int\limits_{t_0}^{t} \! \dd t^\prime \int\limits_{t_0}^{t^\prime} \! \dd t^{\prime\prime} \; \Gamma^{\mu}_{\;\alpha \beta}\left[x_{(0)}^\rho(t^{\prime\prime})\right] v^\alpha v^\beta. 
\end{split}
\end{align}
For clocks that are only accelerated by the GW but otherwise being at rest, we have $v^0 = c$ and $v^{i} = 0$.
In addition, we consider only the $z$ component of $ x_T^\mu$, i.\,e.  $x_T^z(t)=z_T(t)$, and use $\Gamma^z_{\; 00}(z_0,t) = \left.\frac{1}{2}\partial_z h_{00}\right|_{z_0} = h_{+} k_\text{G}^2 z_0 \cos \Phi_t/2$. 
Thus, we obtain
\begin{align}
\begin{split}
    \frac{z_T(t)}{z_0} =&\frac{1}{2}h_{+} \left[ \cos \Phi_t - \cos \Phi_{t_0}+\omega_\text{G}(t-t_0) \sin \Phi_{t_0}\right] 
\end{split}
\end{align}
for the classical trajectory of the trap moving on a geodesic.
Provided that the atom is initially in the ground state of the trap and since Newtonian gravity is neglected, all the dynamics of the atom are encoded in the trap center, i.\,e., $z_T(t)=z(t)$, in accordance with the equivalence principle.
As discussed in Sec.~\ref{sec:Principles}, this result leads to a leading-order sensitivity but requires very challenging setups in terrestrial experiments, while being well-suited for space missions.

\section{Phases for composite interrogation protocols}
\label{app.composite_pulses}
Composite pulse protocols can be used to enhance the sensitivity of atomic sensors, also to GWs.
These protocols usually operate through multiple interactions of the atom with laser pulses.
For atom interferometers, due to the (necessary) momentum recoil, these protocols give rise to large-momentum-transfer techniques.~\cite{Graham2013,Rudolph2020,Gebbe2021}
For clocks, where the recoil is usually suppressed, hyper-Ramsey or echo protocols have been proposed.~\cite{Yudin2010,Zanon2015,Hobson2016,Norcia2017,Zanon2025}
Here, we propose a scheme using exactly the same sequence of $4N-1$ pulses for both, shown in Fig.~\ref{fig:GradSequ}, with the interaction time $t_\ell^{(a/b)}$ of pulse $\ell$ of interferometers $a$ and $b$.

With $\tau = L/c$, we obtain for the first beam-splitter sequence $1 \le \ell \le N$ the times $t_\ell^b/\tau = 2\lfloor \ell/2 \rfloor$ and $t_\ell^a/\tau = 2\lfloor (\ell-1)/2 \rfloor + 1$.
Accordingly, for the mirror sequence $N+1 \le \ell \le 3N-1$ we have $(t_\ell^b -T)/\tau =  2\lfloor (\ell+1)/2 \rfloor-2N$ and $(t_\ell^a -T)/\tau =  2\lfloor \ell/2 \rfloor-2N+1$.
Finally, for the second beam splitter $3N \le \ell \le 4N-1$, we find $(t_\ell^b -2T)/\tau =  2\lfloor \ell/2 \rfloor-4N+2$ and $(t_\ell^a -2T)/\tau =  2\lfloor (\ell+1)/2 \rfloor-4N+1$.

\subsection{Large-momentum-transfer Mach--Zehnder scheme}
To account for the momentum recoil in light-pulse atom interferometers correctly, we choose for the first and second beam splitters, i.\,e.,  for the pulses $1 \le \ell \le N$ and $3N \le \ell \le 4N-1$, the state response functions $\lambda_{k,\ell}=1$ and $\lambda_{\omega,\ell}=(-1)^{\ell+1}$.
For the mirror pulse we have $\lambda_{k,2N} = -2 =\lambda_{\omega,2N}$, while for all other pulses of the mirror sequence, i.\,e., $\ell \in \{N+1, ... , 3N-1\} \setminus\{ 2N\}$, we find
$\lambda_{k,\ell}=-1$ and $\lambda_{\omega,\ell}=(-1)^{\ell+1}$.
With this choice and the timings from above, we find the four differential-phase contributions
\begin{widetext}
\begin{subequations}
\begin{align}
    \delta \phi_\text{G}^\text{AI} &= 2 h_{+} k_\text{L}L \sin \frac{\omega_\text{G}T}{2}\sin \frac{N k_\text{G}L}{2}   \,
    \frac{ \sin \Phi_{(T+N \tau)/2}- \cos  k_\text{G}L  \sin\Phi_{(3T-(N-2)\tau)/2}}{\sin  k_\text{G}L }, \\
    \delta \phi_{\Delta m, \text{G}}^\text{AI} &= -h_{+} (k_\text{G}L)^2 \frac{k_\text{A}}{k_\text{G}} \sin \frac{\omega_\text{G}T}{2}   \sin  \frac{N k_\text{G}L}{2}   \sin\Phi_{(3T-(N-2)\tau)/2},\\
    \delta \phi_{\omega, \text{G}}^\text{AI} &= -4h_{+} \frac{k_\text{L}}{k_\text{G}} \sin \frac{\omega_\text{G}T}{2} \sin  \frac{N k_\text{G}L}{2} \sin \frac{\omega_\text{G}T-(N-1) k_\text{G}L }{2}   \cos \Phi_{T+\tau/2}- \frac{k_\text{L}}{k_\text{A}}  \delta \phi_{\Delta m, \text{G}}^\text{AI}, \\
    \delta \phi_{k, \text{G}}^\text{AI} &= - \delta \phi_\text{G}^\text{AI}.
\end{align}
\end{subequations}
\end{widetext}
Under the approximation $k_\text{L} = k_\text{A}$, where corrections are of higher order and can be incorporated together with finite speed of light,~\cite{Tan2016,DiPumpo2023,Niehof2025} we find that these contributions correspond to Einstein, Doppler, and Shapiro phases.~\cite{Badurina2025} 
Specifically, the term $\delta \phi_\text{G}^\text{AI}$ corresponds to the Doppler shift, while the two terms $\delta \phi_{\omega, \text{G}}^\text{AI}$ and $\delta \phi_{k, \text{G}}^\text{AI}$ represent the Shapiro shift, and $\delta \phi_{\Delta m, \text{G}}^\text{AI}$ is the Einstein shift.
Hence, their sum yields
\begin{align}
\begin{split}
    \delta \phi^\text{AI} =& -2 h_{+}N k_\text{L}L \sin  \frac{\omega_\text{G}T}{2} \sinc  \frac{N k_\text{G}L}{2}\\
    &\times\sin  \frac{\omega_\text{G}T - (N-1)k_\text{G}L}{2}     \cos \Phi_{T+\tau/2}.
\end{split}
\end{align}
For $N k_\text{G} L \ll 1$, we get the signal amplitude 
\begin{equation}
    \phi_\text{S} = \sqrt{2}h_{+} N k_\text{L} L \sin^2  \frac{\omega_\text{G}T}{2},
\end{equation}
which is used in Eq.~\eqref{eq:FinalResMZIAI}.

\subsection{Hyper-echo clock configuration}
Although the pulse sequence for hyper-echo configurations is in principle the same as that for large-momentum-transfer Mach--Zehnder atom interferometers, the effects of the pulses differ for clocks.
As a consequence of trapping (or using recoilless pulses), always both interfering components are addressed by a pulse due to an absence of Doppler detuning.
This difference is reflected by modified values for $\lambda_{k,\ell}$ and $\lambda_{\omega,\ell}$.

As such, for the first beam splitter, we obtain $\lambda_{k,1}=1=\lambda_{\omega,1}$ and for the other pulses $\lambda_{k,\ell}=2$, $\lambda_{\omega,\ell}=(-1)^{\ell+1} 2 $ for $2 \le \ell \le N$, where the latter is also true for the second beam splitter in the interval $3N \le \ell \le 4N-2 $, while $\lambda_{k,4N-1}=1=\lambda_{\omega,4N-1}$.
For the mirror pulse,  we find $\lambda_{k,\ell}=- 2$ and $\lambda_{\omega,\ell}=(-1)^{\ell+1} 2 $ for $N+1 \le \ell \le 3N-1$.
Together with the timings from above, we arrive at
\begin{subequations}
\begin{align}
    \delta \phi_k^\text{C} &= 2 \delta \phi_\text{G}^\text{AI} +2h_{+}k_\text{L}L \sin^2  \frac{\omega_\text{G}T}{2} \cos \Phi_{T+\tau},\\
    \delta \phi_{\Delta m, \text{G}}^\text{C} &= 2 \delta \phi_{\Delta m, \text{G}}^\text{AI} + h_{+}k_\text{A}k_\text{G}L^2 \sin^2  \frac{\omega_\text{G}T}{2}  \sin\Phi_{T+\tau},\\
\begin{split}
    \delta \phi_{\omega, \text{G}}^\text{C} &= 2 \delta \phi_{\omega, \text{G}}^\text{AI} + h_{+}k_\text{L}L \sin^2  \frac{\omega_\text{G}T}{2} \\
    & \times \left(  2\sinc  \frac{k_\text{G}L}{2}  \cos\Phi_{T+\tau/2} -k_\text{G}L \sin \Phi_{T+\tau} \right), 
\end{split}\\
\delta \phi_{k, \text{G}}^\text{C} &= - \delta \phi_k^\text{C}.
\end{align}
\end{subequations}
Adding all these differential phases yields
\begin{align}
\begin{split}
    \delta \phi^\text{C} &=2 \delta \phi^\text{AI}  + 2 h_{+} k_\text{L} L \sin^2  \frac{\omega_\text{G}T}{2} \sinc \frac{k_\text{G}L}{2} \cos\Phi_{T+\tau/2}
\end{split}
\end{align}
for clocks.
In the limit $N k_\text{G} L \ll 1$, we arrive at the signal amplitude 
\begin{equation}
    \phi_\text{S} = \sqrt{2}h_{+} (2N-1) k_\text{L} L \sin^2 \frac{\omega_\text{G}T}{2},
\end{equation}
used in Eq.~\eqref{eq:HyperSignAmp}.

\section*{References}
\bibliography{Literatur}

\end{document}